\documentclass[10pt, conference]{IEEEtran}

\usepackage{multirow}
\usepackage[utf8]{inputenc}
\usepackage{graphicx}
\usepackage[font=footnotesize]{caption}
\usepackage{amsmath}
\usepackage[T1]{fontenc}
\usepackage{subfigure}
\usepackage[english]{babel}
\usepackage[colorlinks]{hyperref}
\usepackage{amsmath,amsfonts}
\usepackage{cite}
\usepackage{graphicx}
\usepackage{epstopdf} 

\usepackage[normalem]{ulem}

\usepackage[vlined,algoruled,linesnumbered]{algorithm2e}
\usepackage[noend]{algpseudocode}
\usepackage[dvipsnames,svgnames,x11names]{xcolor}
\usepackage[markup=underlined]{changes}
\usepackage{graphics}

\graphicspath{{.}{img/}{fig/}{gpi/}}
\def\BibTeX{{\rm B\kern-.05em{\sc i\kern-.025em b}\kern-.08em
    T\kern-.1667em\lower.7ex\hbox{E}\kern-.125emX}}

\begin{document}
\graphicspath{{.}{img/}{fig/}{gpi/}}
 \bstctlcite{IEEEexample:BSTcontrol}
\title{An Attack Resilient PUF-based Authentication Mechanism for Distributed Systems
}
\pagestyle{plain}

\definechangesauthor[color=BrickRed]{NK}
\definechangesauthor[color=NavyBlue]{MY}
\definechangesauthor[color=Green]{ME}
\definechangesauthor[color=Orange]{WL}

\setcommentmarkup{\todo[color={authorcolor!20},size=\scriptsize]{#3: #1}}

\newcommand{\note}[2][]{\added[#1,comment={#2}]{}}

\author{\vspace{0em} \IEEEauthorblockN{Mohammad Ebrahimabadi,
Mohamed Younis, Wassila Lalouani, and Naghmeh Karimi}
\IEEEauthorblockA{CSEE Department, 
University of Maryland Baltimore County, Baltimore, MD 21250\\
}}


\pagestyle{plain}
\maketitle

\begin{abstract}\label{sec:abs}
In most PUF-based authentication schemes, a central server is usually engaged to verify the response of the device’s PUF to challenge bit-streams. 
However, the server availability may be intermittent in practice. 
To tackle such an issue, this paper proposes a new protocol for supporting distributed authentication while avoiding vulnerability to information leakage where CRPs could be retrieved from hacked devices and collectively used to model the PUF.
The main idea is to provision for scrambling the challenge bit-stream in a way that is dependent on the verifier. The scrambling pattern varies per authentication round for each device and independently across devices. In essence, the scrambling function becomes node- and packet-specific and the response received by two verifiers of one device for the same challenge bit-stream could vary. 
Thus, neither the scrambling function can be reverted, nor the PUF can be modeled even by a collusive 
set of malicious nodes. 
The validation results using data of an FPGA-based implementation 
demonstrate the effectiveness of our approach in thwarting PUF modeling attacks by collusive actors.  We also discuss the approach  resiliency  against impersonation, Sybil, and reverse engineering attacks.
\end{abstract}
 
\vspace{-0.05in}
\section{Introduction}\label{sec:intro}
\vspace{-0.05in}
The Internet of Things (IoT) refers to interconnecting miniaturized devices at a large scale, to serve many application domains such as smart transportation, home automation, power grid, and digital battlefield~\cite{YOUNIS201866}.
However, the scale, heterogeneity, ad-hoc topology formation, and dynamic interaction of the connected devices 
make IoT security a major challenge. If left unguarded, the network could be joined with malicious nodes that can apply a wide variety of attacks, such as leaking sensitive data, introducing black hole in the routing topology, and impersonating nodes~\cite{wan-20-internet}.  
Hence, device authentication is highly crucial. The Public Key Infrastructure~(PKI) and Identity-Based Encryption~(IBE) schemes~\cite{Chatterjee-19-Building} have traditionally been used for authentication purposes. However, their high overhead makes them unfit for IoT devices. 

Lightweight hardware security primitives, and in particular Physically Unclonable Functions (PUFs) have received a lot of attention in recent years as an alternative to PKI and IBE schemes~\cite{Chatterjee-19-Building}. Thanks to the unintentional process variations accruing during the manufacturing of integrated circuits, a PUF generates a unique signature that corresponds to its input and output pairs, so-called Challenge-Response Pairs (CRPs).  A PUF is embedded in each device during the fabrication, and a subset of its CRPs are registered after the device fabrication and during the enrollment phase. These CRPs are then used during operation to authenticate the device~\cite{aysu-15-end}. Most existing PUF-based authentication protocols, e.g., ~\cite{SHAMSOSHOARA2020107593},  rely on a central server, which
is not always practical due to  
intermittent connectivity 
or occasional server unavailability. 

To fill the technical gap, this paper promotes a novel Distributed Authentication Using PUFs (DAUP). In DAUP, a pair of IoT devices can mutually authenticate each other through direct exchange of CRPs. Each device stores a small subset of the CRPs of other nodes in the IoT framework. DAUP addresses two fundamental issues related to the use of CRPs in distributed systems: 1) How many CRPs a node $N_i$ should share with any other IoT node? Indeed, there is a tradeoff between achieved security and storage overhead in each node; 
2) How to counter the threat of CRPs sharing among compromised nodes. 
To elaborate, assume that a node stores $M$ CRPs for each other node in the IoT framework. We refer to the set of CRPs of node $N_i$ stored in $N_j$ as $S_{i,j}$.
A question would be how to select these sets, particularly, how different $S_{i,j}$ and $S_{i,k}$ should be, e.g., disjoint, intersecting,\,or completely similar. 
If $S_{i,j}$=$S_{i,k}$, an adversary who captures\,one of these nodes (say $N_j$) can impersonate $N_i$ when interacting with $N_k$.  
On the other hand, storing completely dissimilar CRPs for $N_i$ on other nodes, i.e., $S_{i,j}\cap S_{i,k}=\phi$,
increases the vulnerability to the PUF modeling attack in case $N_j$ and $N_k$ collude and share the CRPs of $N_i$ that they are aware of. 

DAUP employs a novel challenge-bits scrambling scheme that is a function of the node identifier and also changes per packet. A node $N_i$ determines the response for a challenge by factoring in the actual PUF output and the verifier identity
, yet verifiers $N_j$ and $N_k$ would expect different responses for the same challenge given to $N_i$, and thus $N_j$ and $N_k$ fail in modeling the PUF of $N_i$  even if they collude to do so. DAUP is applicable to the widely-used arbiter PUF and its derivatives and results in increased  protection against modeling attacks even through collusion of multiple malicious nodes. 
In summary, the paper makes the following contributions: 
 
\begin{itemize}
\item Devising an effective PUF-based distributed authentication protocol while safeguarding against PUF modeling attacks through collusive group of malicious nodes; 
\item Studying the impact of scrambling the challenge bit-stream on the success of the PUF modeling attacks launched via state-of-the-art ML techniques;
\item Developing an approach for node- and packet-specific challenge scrambling to thwart collusion attacks that opt to model the embedded PUF;
\item Analyzing the resiliency of DAUP against different attack scenarios, in particular cases where the adversary eavesdrops on communication links to intercept  transmissions.
\item Evaluating the proposed method using the data extracted from FPGA implementation of the target PUF.
\end{itemize}

\emph{Note that the novelty of our work is in the distributed aspect of the PUF-based authentication process while countering vulnerability to PUF modeling attacks even when multiple malicious nodes collude.}



\vspace{-0.1cm}
\section{Related work}\label{sec:related work}

Although supposed to be unclonable, a PUF's behavior may be modeled using Machine Learning (ML) schemes~\cite{ebrahimabadi-21-vlsid}. To tackle modeling attacks hardware- and protocol-based methods have been proposed. The former mainly introduces additional circuits, e.g.,~\cite{Zalivaka2019}. However, such a strategy imposes significant area, power, or traffic overhead (e.g.~\cite{gu-19-modeling}), or the provisioned protection can be voided if one node is captured and the function is revealed since the same function is used for all nodes (e.g.,~\cite{Zalivaka2019}).
Protocol-based methods can be classified into: (i) CRP obfuscation, (ii) controlled challenge bit-streams, (iii) noise injection.  CRP obfuscation is applied either through encryption~\cite{Gope2019} or challenge bit-shuffling~\cite{PUF-RAKE}. The latter can also be realized through challenge partitioning over multiple packets~\cite{Ebrahimabadi-21-PUF}.  However, these methods are only applicable where a central controller is available.


The second category of the protocol-based schemes controls the used challenge bit-streams as a means to deprive an eavesdropper from collecting CRPs for modeling the PUF~\cite{BARBARESCHI2019}. Yet, this category is only applicable when authentication is not conducted very often and their utility in IoT is questionable. Also, a central server needs to be engaged. Finally, adversarial ML has been used to introduce noisy data that degrades PUF modeling attempts~\cite{wang-19-adversarial}~\cite{ebrahimabadi-21-adversarial}. Wang et al.~\cite{wang-19-adversarial} proposed to poison the PUF's response based on the challenge bits. However, such a protection has been defeated by Ebrahimabadi et al.~\cite{ebrahimabadi-21-adversarial}. Although successful against modeling attacks, the approach of~\cite{ebrahimabadi-21-adversarial} imposes high computational overhead due to conducting ML-based modeling quite frequently.

Other PUF-based authentication schemes, e.g.~\cite{lao-16-reliable}, are vulnerable to security threats such as modeling, replay, and impersonation attacks. To alleviate such vulnerability, the use of a cryptosystem along with PUF has been pursued, where the PUF is leveraged to generate keys for securing the data exchange \cite{Chatterjee-19-Building,SRAM-PUF2020}. Despite its effectiveness, such an approach imposes significant overhead on IoT devices. 
Overall, existing PUF-based authentication schemes rely on a central server. This paper enables distributed authentication while safeguarding against PUF modeling attacks by a collusive group of malicious nodes.  
\vspace{-0.1cm}
\section{System Model and Preliminaries}\label{sec:systemmodel}
\subsection{Arbiter-PUF Architecture
}\label{sec:preliminary}

\begin{figure}[h]
  \centering
  \vspace{-1.1em}
   \includegraphics[width=2.5in]{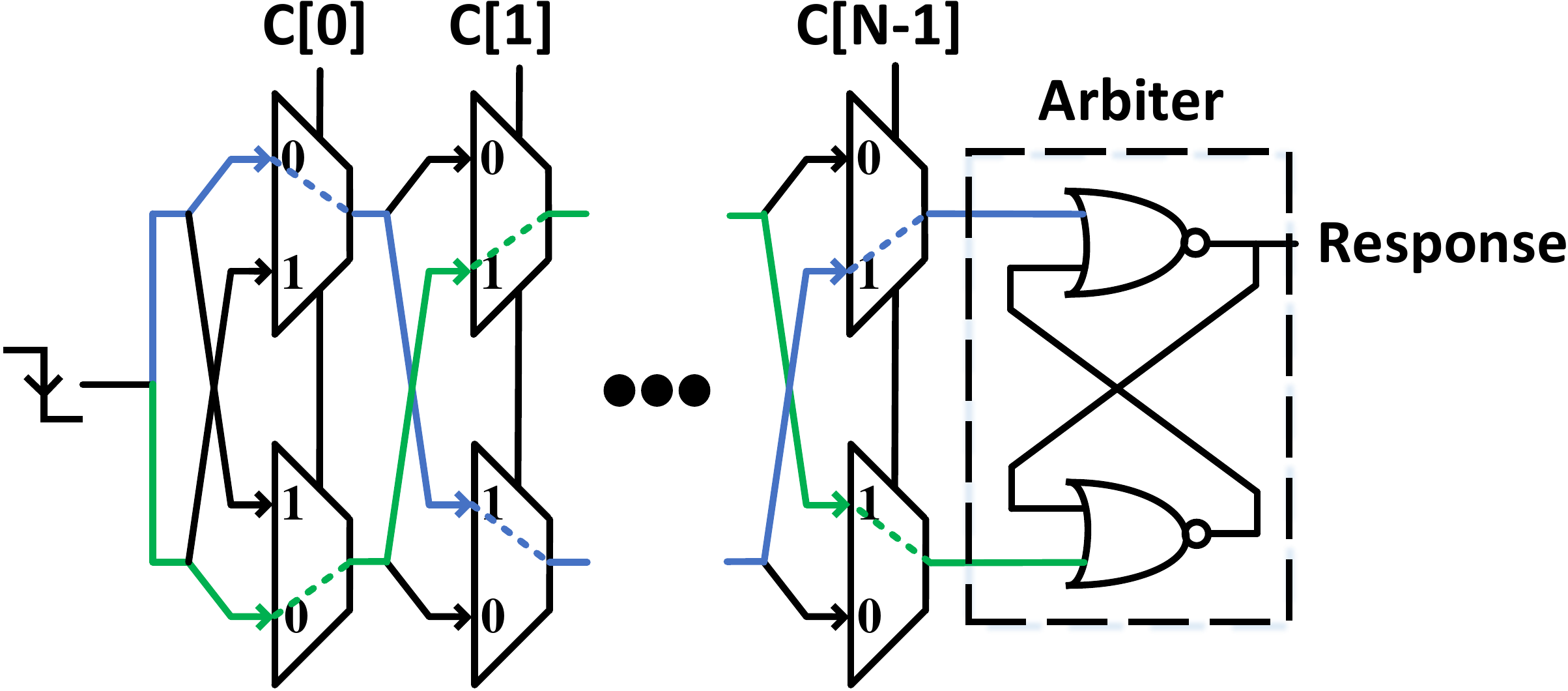}
   \caption{Illustrating the design of an arbiter-PUF.}\label{fig:ARBITER}
   \vspace{-0.4cm}
\end{figure}

The arbiter-PUF is characterized by its low implementation overhead and large CRP count; hence it is deemed effective  
for supporting device authentication. This PUF operates based on manufacturing process variations that result in a race between two signal paths (e.g., blue and green paths shown in Fig.~\ref{fig:ARBITER}), to generate a response bit for a challenge bit-stream~\cite{becker-15-gap}. The race corresponds to difference in propagation delays over these two paths, and affects the value latched by the arbiter circuit. Indeed, the arbiter output (PUF response) is  affected only by the sign of the delay difference and not its amount. 
Although the delays vary based on the queried challenge, and the corresponding response is unpredictable, 
the interception of CRPs can make the arbiter-PUF (as well as its derivatives, e.g., XOR PUF) vulnerable to a modeling attack, where the adversary deploys ML to mimic the PUF behavior. 

\vspace{-0.1cm}
\subsection{System and Threat Models} \label{subsec:threatmodel}


We assume that an arbiter-PUF or one of its derivatives is embedded in each IoT device during fabrication, to be used in identifying the device. To authenticate a target node $N_i$ (so called prover) by node $N_j$ (so called verifier), the latter sends a challenge bit-stream to $N_i$. Then $N_i$ applies the challenge to its PUF and sends the PUF response to $N_j$. 
Upon the receipt of the PUF response, $N_j$ compares it with a pre-known value to confirm the identity of $N_i$. In DAUP, each node can play the role of a verifier and hence should have a subset of CRPs of all other nodes it interacts with. This is done during the enrollment phase, when a device joins the IoT framework. 



In order to collect sufficient CRPs to model the target PUF and impersonate the corresponding device $N_i$, an adversary is assumed to either eavesdrop on the wireless links between $N_i$ and other nodes (verifiers) in the system or hack some of these verifiers and read the stored CRPs of $N_i$ in their memory. We categorize such a threat as collusion since it involves multiple nodes. Our proposed method thwarts such a collusive attack. 


\vspace{-0.1cm}
\section{DAUP Authentication Scheme}\label{sec:method}

DAUP opts to enable PUF-based device authentication without engaging a server as an intermediary, and while countering the threat of PUF modeling by: (i) an eavesdropper on the communication links, and (ii) hacking (intruding) to one or multiple nodes.   A server is  involved only during the enrolment of a node $N_i$ to securely provide some shared CRPs between $N_i$ and other nodes. 
The shared CRPs could be updated, e.g., to prevent replay attack, when the server is reachable. 
This section describes DAUP in detail.

\subsection{Detailed Design}\label{sec:Design}

\noindent
\underline{Support for Distributed Authentication:} 
Similar to other PUF-based authentication schemes, DAUP assumes that a PUF is embedded in each IoT device in the fabrication process. During the system initialization phase, when a device $N_i$ is enrolled in an IoT framework, a set of challenge bit-streams, ${\Gamma}^i$, are generated and given to the device's embedded PUF, and their related responses are tabulated to be used for authenticating $N_i$ after field deployment. We note that in DAUP the response of a challenge bit-stream depends on the verifier’s ID where each challenge in ${\Gamma}^i$  will be subject to scrambling before applying to the PUF of $N_i$, as we explain below; hence the server will store for $N_i$ the sets $CRP_{i,j}$ corresponding to ${\Gamma}^i$ for $\forall$ $N_j$ in the system where $ j \ne i$. \emph{We note that the initialization phase would engage a server where the CRPs for all enrolled devices are saved. Such a server, which is assumed to be trusted, is not involved in the mutual authentication of IoT nodes}; this 
prevents the server from becoming a bottleneck, especially when the communication links among nodes are intermittent and frequent authentication is necessary. We also note that in DAUP the server will not store the actual response of PUF of $N_i$ for ${\Gamma}^i$ during the enrolment, instead, it only stores the response of a  scrambled version of ${\Gamma}^i$, and hence the server does not constitute a modeling threat. 

Each device $N_j$ will receive from the server a subset $crp_{i, j}$ of $CRP_{i, j}$ $\forall$ $ j \ne i$. The cardinality of $crp_{i, j}$ is usually subject to tradeoff. On the one hand, having many CRPs in $crp_{i,j}$ allows a device $N_j$ to switch among multiple challenges over time and thus increases the robustness of the authentication process. On the other hand, the aggregate memory size needed for $N_j$ for storing $crp_{i, j}$ $\forall$ $ j \ne i$ constitutes overhead and minimization of such overhead could be desirable to cope with resource constraints, especially for a large system with many nodes. We expect, nonetheless, that the lifetime of the system will play a dominant role since the required storage space is not much by today’s standard, e.g., for a PUF with 64-bit challenge and 32-bit response, and a 100-node system, each node will need about 120K bytes for storing 100 CPRs for all other devices in the system. Note that to save power consumption, a multi-bit response is usually extracted by querying the PUF multiple times using the challenges generated internally based on the given challenge~\cite{PUF-RAKE}. Indeed, the susceptibility to modeling attacks is not a factor in determining the size of $crp_{i, j}$ since challenge bit scrambling proves to safeguard the individual PUFs against an adversary that intercepts challenge response exchange among nodes or even hacks a verifier to read its memory, as we discuss below. Finally, we would like to stress that the initialization phase is a byproduct of the distributed operation of an IoT rather than something that DAUP dictates. Particularly, how to provide $crp_{i, j}$ to nodes when the system is set up and how to handle node addition and departure (e.g., when a node fails) is a general issue for  distributed operation and not particularly imposed by how authentication is conducted. 

\vspace{3pt}
\noindent
\underline{Underlying PUF Modeling Countermeasure:}
PUFs, and in particular arbiter-PUFs, are vulnerable to ML-based modeling attacks, i.e., by accessing some CRPs an adversary can predict the PUF response of an unseen challenge bit-stream. Our protection strategy against such type of attacks is through scrambling the challenge bit-stream, $C$, in order to de-correlate $C$ from the PUF response as seen by an eavesdropper.
The basic idea is to reorder the challenge bits before applying it to the PUF~\cite{Ebrahimabadi-21-PUF}. For example, instead of applying the challenge bit-stream received from the verifier as $C[0], C[1], ..., C[N-1]$ in Fig.~\ref{fig:ARBITER}, a shuffled version like ${C[15], C[N-4], ..., C[40]}$ is given to the target PUF. Such reordering misleads the eavesdropper's ML model, as the intercepted response $R$ would be for the Scrambled Challenge (SC) and not for $C$ itself. Note that in DAUP the scrambling pattern is different per challenge and also per verifier node. As will be discussed in Section~\ref{sec:exp}, the adversary is unable to brute-force DAUP and uncover the unscrambled challenges. 


\vspace{3pt}
\noindent
\underline{Thwarting Collusive Attacks:} 
By scrambling the challenge bit-stream, DAUP tackles modeling attacks launched by an adversary who captures CRPs of the targeted node, $N_i$, through eavesdropping on $N_i$ communications or hacking and reading the memory of  verifiers that deal with $N_i$. As noted, such mitigation is sufficient to protect an individual node as long as the adversary does not know how to unscramble the bit-steam, as otherwise by intercepting enough CRPs the adversary could model the PUF~\cite{Ebrahimabadi-21-PUF}. 
Meanwhile, if a similar scrambling pattern is used by multiple verifiers, hacking one verifier or intercepting the CRPs it uses for a node $N_i$,  will allow the response of $N_i$ to be known for certain challenges and does not prevent impersonation. 
Worse, if the scrambling function is fixed, hacking a node will make the entire network vulnerable. 

To alleviate such vulnerability, DAUP considers a scrambling function that varies per verifier and per challenge. Basically, node $N_i$ will scramble the challenge bit-stream $C$ from verifier $N_j$ based on the $ID_j$ and the value of $C$ itself. Hence, not only the response for the same challenge at different verifiers will be different, but also the scrambling pattern for challenges sent by the same verifier is different. This is a very powerful feature as conflicting data will be fed to the adversary's ML model which   degrades its accuracy. To illustrate, let $R=F_i(C)$ be the PUF function for node $N_i$. According to DAUP,  a verifier $N_j$ will have $\hat{R}=F_i(\zeta_{i,j}(C))$, where $\zeta_{i,j}$ is the scrambling function that $N_i$ applies for $N_j$, when $N_j$ authenticates $N_i$ with $C$. Generally $\hat{R} \ne R$; hence $N_j$  tabulates ($C,\hat{R}$)  to be used for authenticating $N_i$.  DAUP benefits from the PUF not only in device authentication but also to devise the challenge scrambling pattern, as explained in the next subsection. Thus the scrambling pattern is both node-specific and challenge-specific, and differs from one node to another and one challenge to another. Indeed, this is one of the DAUP's main advantages.

\vspace{-0.1cm}
\subsection{DAUP Implementation}\label{sec:CSP-S}
Fig.~\ref{fig:scrambling} shows a block diagram description of DAUP, where the steps are grouped into two sets: (1) those implemented in software and mainly reflect the communication interface, and (2) steps realized in hardware and correspond to the underpinning tamper-proof protection. The software part of DAUP deals with receiving  requests from verifiers;  a request from a verifier $N_j$ will include its ID, which along with the queried challenge, $C$, 
are used to determine the scrambling function (pattern). Such a pattern is generated in hardware (bottom block), and is applied to $C$ on the fly to generate the corresponding response that will be sent back to $N_j$. This response will then be checked against the value stored during the enrolment for such a challenge in $N_j$. 
DAUP operates in two phases: 
\begin{figure}[htb]
 \vspace{-1em}
  \centering
   \includegraphics[width=3.0in]{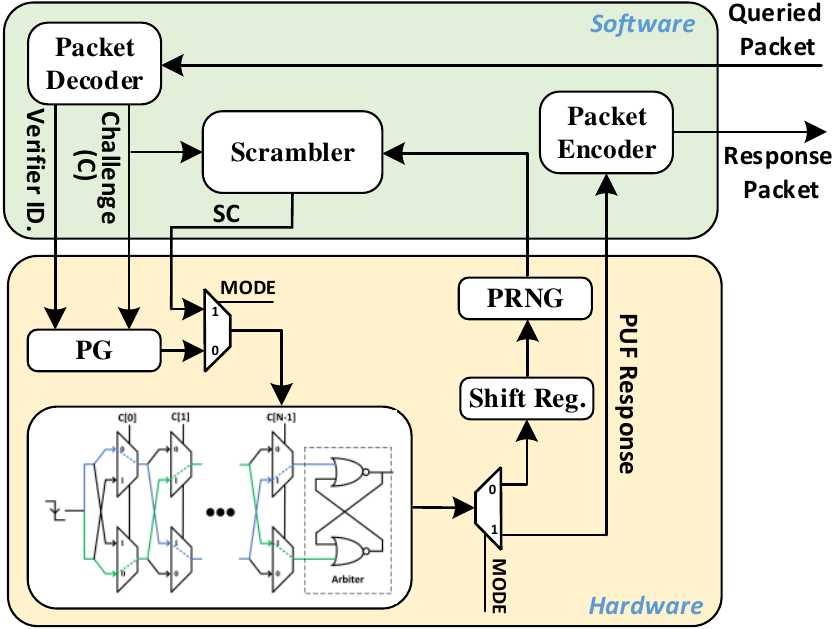}
   \caption{DAUP Block Diagram. 
   }\label{fig:scrambling}
   \vspace{-1em}
\end{figure}

\noindent\emph{1) Determining a scrambling pattern}: When node $N_i$ is queried by $N_j$ with a challenge bit-stream $C$, DAUP goes to phase~0 ($MODE=0$) to 
    define how $C$ will be transformed. The mode signal controls the multiplexer (MUx), and de-multiplexer (de-Mux). The PUF is engaged in generating the scrambling pattern as we explain later;  
    
\noindent\emph{2) Generating response for a verifier's request}: In this phase, the picked scrambling pattern is used to reorder $C$ and generate $SC$ which in turn is fed to the PUF. The PUF response is then sent to $N_j$ to be used for authenticating $N_i$. By setting ($MODE=1$), the Mux. (and de-Mux.) will allow $SC$ to be a PUF input and the corresponding response to be directed to the packet encoder within the software part of DAUP. 



%
Alg.~\ref{alg-auth} summarizes the steps taken in DAUP when a verifier $N_j$ authenticates $N_i$. 
After extracting the challenge (C) and the verifier ID, $ID_j$, from the request (line~\ref{line1}), the  $MODE$ is set to ``0'' to determine the scrambling pattern being applied to $C$. 
An LFSR-based Pseudo Random Number Generator (PRNG) is employed to devise the scrambling pattern. To make DAUP more resilient against modeling attacks as well as brute forcing of the scrambling patterns, for each challenge, a different seed is used to initialize the embedded PRNG based on the verifier's ID as well as the value of the given challenge itself (lines~\ref{line3}-\ref{line9}).

\vspace{-0.8em}
\DontPrintSemicolon
\begin{algorithm}[ht]
	\small
	\SetKwInOut{Input}{input}
	\SetKwInOut{Output}{output}
	\Input{Queried packet received from the verifier node $N_j$}
	\Output{Response packet to be sent to the verifier $N_j$}
	\BlankLine
	Decode the queried packet to extract Challenge (C) and the verifier ID
	($ID_j$)\label{alg-line1} \;\label{line1}
	$MODE \gets 0$\;\label{line2}
	
	
	$MC_1 \gets C[0\!:\!N\!-\!S\!-\!1] \ || \ ID_j$\;\label{line3}
	\tcp*[r]{\scriptsize S: bit-length of $ID_j$, N: PUF Challenge-length}
	$K \gets log_2N$\;\label{line4}
	
	\For{$h\in\{1,\ldots, K\}$}{\label{line5}
    	$R_h \gets PUF(MC_h)$\;\label{line6}
    	$Shift \ Reg.[K-1:0] \gets Shift\ Reg.[K-2:0] \ || \ R_h$\;\label{line7}
    	$MC_h \gets MC_h>>1$\;\label{line8}
	}
    $PRNG(Init) \gets Shift \ Reg.$\;\label{line9}
    $SC[0] \gets C[0]$\;\label{line10}
	\For{$h\in\{1,\ldots, N-1\}$}{\label{line11}
	$H_h \gets PRNG(h)$\;\label{line12}
	$SC[h] \gets C[H_h]$\;\label{line13}
	}	
	$MODE \gets 1$\;\label{line14}
	$R \gets PUF(SC)$\;\label{line15}
	\Return Response packet that includes $R$\;\label{line17}
	\caption{DAUP authenticating node $N_i$\,by $N_j$}
	\label{alg-auth}
\end{algorithm}
\vspace{-0.8em}

Assume that the challenge length, $N$, is larger than $ID_j$ bit-length ($S$), and the initial Mutated Challenge ($MC_1$) is built by concatenation (shown as || in line~\ref{line3}) of $ID_j$ and $N-S$ bits of the received challenge. We feed the PUF with $MC_1$, and push the related PUF response to a shift register that builds the PRNG seed bits gradually. Indeed as shown in lines~\ref{line5}-\ref{line8}, the Pattern Generator (shown as PG in Fig.~\ref{fig:scrambling}) builds the bit values of the LFSR's seed, stored in the Shift Register, one by one by feeding the PUF with the last mutated challenge each time. To generate each mutated challenge ($MC_h$) DAUP  circularly shifts the previous mutated challenge, ($MC_{h-1}$) one bit to the right. Note that any other function can be used to generate mutated challenges as well;  due to its low implementation overhead, we have decided to use a circular shift function.

The PRNG is implemented as a $K= log_2 N$ bit LFSR with \emph{primitive polynomial}, 
e.g., for a 64-bit challenge a 6-bit LFSR suffices. By definition, a primitive polynomial will cycle through all possible non-zero states~\cite{ghorpade-11-primitive}. After initializing the LFSR with the $K$ bit PUF response stored in the shift register, we clock the LFSR $N-1$ times. This will result in $N-1$ distinct values, $H_1,H_2, ...,H_{N-1}$, for each of K bits, thanks to the primitive polynomial of LFSR. As shown in lines~\ref{line10}-\ref{line13}, concatenating the generated list with ``0'' will form the ordered list of $[0, H_1,H_2, ...,H_{N-1}]$ (line \ref{line12}), which in turn is used to map the initial challenge bits, $C$, to its scrambled counterpart, $SC$. 
Upon applying $SC$ to  $N_i$'s PUF in phase~1 ($MODE=1$), the corresponding response is sent to the verifier $N_j$ (lines ~\ref{line14}-\ref{line17}), where it is matched with the expected (pre-tabulated) values. 
Even in the rare case of having a PUF response of `0' for all $K$ queries in phase~0 (line~\ref{line6}), and thus initializing the LFSR with a seed equal to zero, DAUP still works. In this case, as the LFSR gets stuck in state zero (i.e., all of its bits are `0'), based on lines\ref{line10}-\ref{line13}, SC[h] would get C[0] for $h\in\{1,.., N-1\}$. In this case $C$ is mutated, instead of scrambled, and the related $SC$ can be used for authentication as the verifier has stored the response for such a mutated challenge.

Given that the LFSR generates $N-1$   (rather than $N$) unique values, DAUP does not change the location of $C[0]$ which is the furthest bit from the arbiter (see Fig.~\ref{fig:ARBITER}), while it maps $C[1],C[2],..,C[N-1]$ to $C[H_1],C[H_2],..,C[H_{N-1}]$, denoted as $SC[1]$ to $SC[N-1]$, respectively. We argue that keeping the value of C[0] intact before and after scrambling does not have a considerable effect, as the further the challenge bit is from the arbiter, the lower the impact it has on the delay chains~\cite{ebrahimabadi-21-vlsid}. 
Note that in case $S$~>~$N$, the initial mutated challenge $MC_1$ is formed of the  $F<N$ (e.g., N/2) least-significant bits of $ID_j$ instead of the whole bit-stream, concatenates with $N-F$ bits of $C$. This constitutes a special case for Line~\ref{line3} and is not shown in the algorithm for the sake of simplicity.




 \vspace{-0.3em}
\section{Experimental Results and Discussions}\label{sec:exp}
We validated DAUP using 6 Xilinx ARTIX7 FPGA boards, each representing an IoT device and assigned a 32 bit unique ID. The nodes are connected using Zigbee transceivers. 
A 64-bit arbiter-PUF and a 6-bit LFSR with the $X^6 + X^5 + 1$ primitive polynomial have been implemented on each device. The latter is to generate the scrambling patterns. 
The adversary is assumed to intercept the exchanged CRPs between the targeted node $N_t$ (i.e., prover) and the verifier nodes $N_v$ where $ 1\le v\le 5$. 
We extracted the response for a set of 22,000 randomly chosen challenges applied to the PUF of $N_t$.


To show the resiliency of DAUP against  modeling attacks that use state-of-the-art ML schemes, we consider the cases where the adversary uses Neural Network (NN), Support Vector Machine (SVM) or Logistic Regression (LR) as the representatives of ML techniques. We used a 5-layer fully connected NN with one input layer (with 64 neurons reflecting the PUF\,size), three nonlinear hidden layers (with 5, 10 and 15 neurons) and one output neuron with a sigmoid function. A rectified linear unit (ReLU) is used as an activation function in every layer. The learning rate, momentum, and \# of epochs are 0.01, 0.99, and 2000 respectively.
 \vspace{-0.2em}
\subsection{Experimental Results}\label{sec:scr}
 \vspace{-0.2em}
 \begin{figure}[b]
  \centering
   \includegraphics[width=0.45\textwidth]{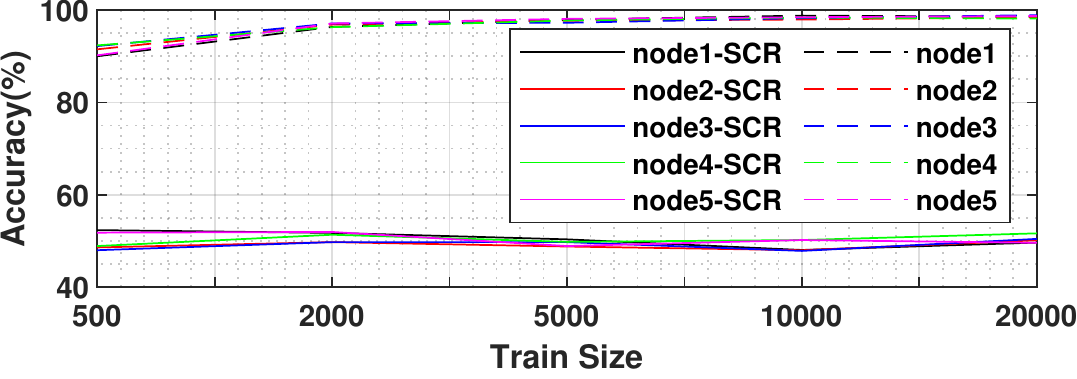}
   \vspace{-0.1em}
   \caption{Accuracy of the adversary's ML model with and without using DAUP. The PUF model is built using NN based on the CRPs exchanged between the verifier ($N_v, v=1,2,...,5$) and prover($N_t$) nodes while the training size is increased to 20K.}\label{fig:SCR_Tr}
   \vspace{-0.3em}
\end{figure}
\noindent\textbf{Challenge Scrambling:} The first set of results, shown in Fig.~\ref{fig:SCR_Tr}, demonstrates the efficiency of DAUP in diminishing the accuracy of the ML-based model that an adversary builds based on  the intercepted CRPs. Here, we assume that the adversary eavesdrops on the communication line between $N_v$ ($v$=1,2,...,5) and the target node (i.e., prover) $N_t$, and trains a NN model to predict the responses of the unseen challenges fed to $N_t$'s embedded PUF. In Fig.~\ref{fig:SCR_Tr}, the solid and dotted curves relate to the cases where challenge-scrambling is and is not performed, respectively. As shown, when there is no protection, with as low as 500 CRPs, the adversary can train a model successfully and achieve an accuracy of $\approx90\%$; such accuracy grows to 99\% with the training size of 3,000 CRPs. However, when DAUP is applied (solid lines), the accuracy stays close to 50\% even with a training size of 20,000 CRPs.  It is important to note that, because of the binary nature of a PUF response, a 50\% accuracy reflects a random guess of the PUF response. Thus,  by benefiting from challenge scrambling, DAUP can successfully prevent PUF modeling attacks.     

To demonstrate the success of DAUP in thwarting collusive PUF modeling attacks, we consider the three following scenarios where the adversary uses NN models: 

\noindent\uline{Scenario I:}
In this case, the adversary eavesdrops on communications between the target device ($N_t$) and verifier $N_i$. The adversary opts to build a model of $N_t$'s embedded PUF based on the intercepted CRPs in order to predict the responses of $N_t$ to another verifier, say $N_j$, where $j \ne i$. The accuracy of such prediction is listed in Table~\ref{tab:tab1} when 100 (shown preceding the parentheses) and 1000 (shown inside parenthesis) CRPs of $N_t$ are intercepted. Here, we assume that a total of 500 (5000 in the second case) 
randomly selected CRPs for $N_t$ are shared with all five versifiers to authenticate $N_t$.
Note that the verifier nodes may or may not have tabulated similar challenges of $N_t$, yet the stored challenge set can have some overlap. The $N_t$'s responses held by each verifier for the same challenge may be different as the verifier's ID affects the scrambling pattern; recall that the verifier stores the response of the scrambled rather than the original version of the challenge. 
 
The results in Table~\ref{tab:tab1} confirm the effectiveness of DAUP. 
Since DAUP uses a different scrambling pattern based on the verifier's ID and challenge bit-stream, the adversary is unsuccessful in predicting the target PUF's response in all cases. Specifically, if the model is built based on the CRPs exchange between $N_t$ and $N_1$ (i.e., i=1), a verifier $N_2$ (i.e., j=2) is unable to predict the responses of the other challenges exchanged between $N_t$ and $N_1$; even with intercepting 1000 CRPs the success rate is 51\%. Recall that the results for the cases where $i=j$ are shown in Fig.~\ref{fig:SCR_Tr}. 




\begin{table}[h]
\centering
\vspace{-0.7em}
\caption{Adversary's gained accuracy in predicting the response of a prover $N_t$ to a verifier $N_j$ when the adversary builds the $N_t$ model based on the CRPs exchanged between $N_t$ and a verifier $N_i$, $1 \le i,j \le 5, i \ne j$ using NN. The numbers preceding and inside parenthesis reflect the modeling accuracy when 100 and 1000 CRPs used for building the model, respectively.
}
\vspace{-0.2em}
\label{tab:tab1}
\scriptsize
\begin{tabular}{|c|l|l|l|l|l|}
\hline
\multirow{3}{*}{
i} & \multicolumn{5}{c|}{j

}\\ \cline{2-6} 
 & \multicolumn{1}{c|}{1} & \multicolumn{1}{c|}{2} & \multicolumn{1}{c|}{3} & \multicolumn{1}{c|}{4} & \multicolumn{1}{c|}{5} \\ \cline{1-6} 
1 & -------- & 51(51) & 42(47) & 49(48) & 52(50) \\ \hline
2 & 55(47) & -------- & 52(50) & 45(49) & 54(48) \\ \hline
3 & 43(49) & 53(49) & -------- & 51(50) & 41(49) \\ \hline
4 & 34(48) & 44(50) & 48(51) & --------& 48(51) \\ \hline
5 & 52(49) & 52(54) & 47(52) & 51(50) & -------- \\ \hline
\end{tabular}
\end{table}

\noindent\uline{Scenario II:}
In this case, the adversary is able to eavesdrop on multiple communication links, specifically, between verifiers $N_i$ and $N_k$, and the prover $N_t$. The adversary used the intercepted CRP between $N_t$ and those two verifiers to model the PUF of $N_t$ and predict the responses of $N_t$ to another verifier, say $N_j$ where $j \ne i, j \ne k$. The prediction accuracy is reported in Table~\ref{tab:tab2} the adversary trains the PUF model using 100 and 1000 CRPs per verifier, i.e., a total of 200 and 2000 CRPS. The results in Table~\ref{tab:tab2} affirm DAUP's ability to counter modeling attacks even if multiple nodes collude or if the CRPs exchanged between them and the prover are intercepted. The modeling accuracy for more than 2 colluding nodes follows a similar trend, but is not shown due to space constraints.

\vspace{-0.3cm}
\begin{table}[h]
\centering
\caption{Adversary's accuracy in predicting the response of a prover $N_t$ to a verifier $N_j$ when using NN to build a model of $N_t$'s PUF based on CRPs exchanged between $N_t$ and both $N_i$ and $N_k$ verifies. The numbers preceding (inside) parentheses depict the modeling accuracy when 100 (1000) CRPs are exchanged between each verifier and $N_t$ used to train the model.}
\vspace{-0.3em}
\label{tab:tab2}
\scriptsize
\begin{tabular}{|c|c|c|c|c|c|}
\hline
\multirow{3}{*}{i,k} & \multicolumn{5}{c|}{j} \\ \cline{2-6} 
 & \multicolumn{1}{c|}{1} & \multicolumn{1}{c|}{2} & \multicolumn{1}{c|}{3} & \multicolumn{1}{c|}{4} & \multicolumn{1}{c|}{5} \\ \cline{1-6} 
1,2   & -------- & -------- & 48(47) & 50(49) & 50(48) \\ \hline
1,3 & -------- & 65(52) & -------- & 48(50) & 58(51) \\ \hline
1,4 & -------- & 51(50) & 46(51) & -------- & 56(49) \\ \hline
1,5 & -------- & 56(50) & 49(50) & 48(50) & -------- \\ \hline
2,3 & 50(49) & -------- & -------- & 47(51) & 48(48) \\ \hline
2,4 & 48(49) & -------- & 65(49) & -------- & 46(49) \\ \hline
2,5 & 52(47) & -------- & 54(51) & 51(50) & -------- \\ \hline
3,4 & 47(49) & 55(51) & -------- & -------- & 47(50) \\ \hline
3,5 & 50(48) & 56(51) & -------- & 42(50) & -------- \\ \hline
4,5 & 57(48) & 50(52) & 41(51) & -------- & -------- \\ \hline
\end{tabular}
\end{table}


\noindent\uline{Scenario III:}
This scenario assumes that the adversary can intercept a portion (L\%) of all communications between IoT nodes and  $N_t$, and builds the model based on the captured CRPs to predict the response of unseen challenges sent to $N_t$ from verifier $N_j$. The modeling accuracy is provided in Table~\ref{tab:tab3}, when L=10\%, 20\%, .., and 50\% of all  exchanged CRPs between each node and $N_t$ are intercepted and used to train the NN model and in turn to predict the responses for unseen challenges. The obtained results indicate that increasing the intercepted percentage of exchanged  CRPs does not have any significant impact in increasing the accuracy of modeling $N_t$, i.e., in all cases the accuracy is$\approx50\%$. This is a clear testimony for the effectiveness of DAUP.

\begin{table}[h]
\centering
\vspace{-1em}
\caption{The success rate for predicting the response of prover $N_t$ to a verifier $N_j$ when the adversary builds a NN model of $N_t$ based on L\% of the CRPs exchanged between $N_t$ and all other verifiers. The numbers preceding (inside) parenthesis depict the modeling accuracy when 100 (1000) CRPs are transferred between each verifier and $N_t$.}
\vspace{-0.3em}
\label{tab:tab3}
\scriptsize
\begin{tabular}{|c|c|c|c|c|c|}
\hline
\multirow{3}{*}{L\%} & \multicolumn{5}{c|}{j
}\\ \cline{2-6} 
 & \multicolumn{1}{c|}{1} & \multicolumn{1}{c|}{2} & \multicolumn{1}{c|}{3} & \multicolumn{1}{c|}{4} & \multicolumn{1}{c|}{5} \\ \cline{1-6} 
10 & 52(48) & 45(53) & 45(49) & 47(54) & 47(47) \\ \hline
20 & 42(49) & 50(49) & 52(52) & 42(51) & 42(49) \\ \hline
30 & 44(46) & 52(51) & 50(48) & 50(48) & 50(47) \\ \hline
40 & 50(47) & 40(49) & 58(50) & 58(51) & 58(49) \\ \hline
50 & 58(46) & 52(52) & 48(46) & 48(48) & 48(51) \\ \hline
\end{tabular}
\end{table}

\begin{figure}[h]
  \centering
   \includegraphics[width=2.8in]{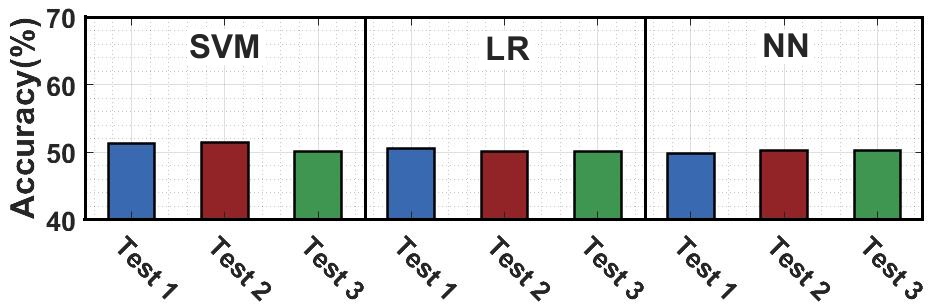}
   \vspace{-0.3em}
   \caption{Modeling accuracy in each of the considered test cases when different ML schemes are used to model the target PUF in presence of DAUP.}\label{fig:ML}
   \vspace{-1em}
\end{figure}

\noindent\textbf{Resiliency Against Various ML Modeling Techniques:}  
This set of results gauge the resiliency of DAUP against modeling attacks that utilize SVM or LR. Due to space limitation, we only consider training and test set sizes of 20K and 2K, respectively. Fig.~\ref{fig:ML} shows the results. Test~1 refers to Scenario I where the adversary eavesdrops on the communication link between $N_1$ and $N_t$ and tries to predict $N_t$'s response when queried by $N_2$. Test~2 realizes the case where the adversary intercepts the exchanged CRPs between $N_t$ and $N_1$ as well as the ones transferred between $N_t$ and $N_2$ (Scenario II). The same results can be obtained if $N_1$ and $N_2$ collude and share the $N_t$ CRPs they have stored. The trained model is then used\,to predict the\,response of $N_t$ to\,the challenges submitted by $N_3$.  Finally in Test~3, we selected\,the last item of scenario III, i.e., the best case scenario for the adversary among those cases where he intercepts 50\% of all communication traffic. The trained model is then used to infer the rest (other 50\%) unseen responses. As Fig.~\ref{fig:ML} shows the modeling accuracy does not exceed 55\% in all these tests\,when NN, SVM, or LR is used.  
The CMA-ES based attack~\cite{becker-15-gap} is not applicable in our case as the response to the same query changes per verifier. 

\vspace{-0.4em}
\section{Overhead and Security Analysis}\label{sec:diss}
\vspace{-0.4em}

\noindent\uline{Protection against Impersonation and Sybil attacks}: 
Impersonation refers to the scenario when a malicious node tries to identify itself as a legitimate node. In DAUP, 
since the challenge bit-streams are scrambled, the attacker is unable to build the prover's PUF model to impersonate it. Also the scrambling pattern changes per verifier and per packet, as well as per prover (due to the use of PUF in the process of generating scrambling patterns). Thus, an adversary cannot infer the scrambling algorithm even by colluding with other malicious nodes. This makes impersonation impossible. Similarly, a Sybil attack where an adversary claims multiple valid identities, is not feasible as an adversary cannot even impersonate a single IoT node. 

\vspace{3pt}
\noindent{\uline{Protection against Reverse Engineering attack}}:
DAUP makes the chip resilient against reverse engineering even if the adversary has access to the IoT device itself and opts to uncover the chip using imaging, delaying, etc. This is because DAUP generates the scrambling pattern for each chip by using the chip's embedded PUF, thus a unique and unpredictable signature, thanks to the process variations. Thereby, even if the algorithm is known, it cannot be compromised. 

\vspace{3pt}
\noindent{\uline{Protection against Brute-Forcing the scrambling pattern}}:
Even if an adversary knows that scrambling is used, he needs to guess the seed value generated based on PUF responses to find the scrambling pattern. Thus for an $N$ bit challenge, the seed value ($=log_2N$ bit) should be guessed. As the seed changes per challenge, for modeling the PUF using $M$ challenges, all possible $N^M$ cases should be checked in order to find the unscrambled pattern and model the PUF.

\vspace{3pt}
\noindent{\uline{Required storage per node}}:
The size of tabulated CRPs in each device (i.e., verifier) depends on the number of devices in the IoT framework ($ND$), authentication rate ($AR$), and  time interval between enrollments ($TI$). The fewer $ND$ is, the more CRPs per node can be stored. 
On the other hand, in case of higher $AR$ and $TI$ rates, more CRPs are tabulated to prevent replay and impersonation attacks. Eq. (\ref{eq:data}) shows the  memory size (in bit) required by DAUP for each IoT device. Here, each CRP includes an $N$-bit challenge and $R$-bit response.
\vspace {-0.05in}
\begin{equation} \label{eq:data}
\small
\begin{split}
&Memory \ Size=  TI\times AR \times (ND-1)  \times (N+R)
\end{split}   
\end{equation}

\vspace {-6pt}

\noindent\uline{DAUP Overhead and noise impact}: DAUP imposes a negligible hardware overhead. The overhead for an IoT device with a 64-bit arbiter-PUF includes: (i) the $PG$ block which performs concatenation and circular shifting operations, being implemented via a 64-bit register and rewiring, (ii) a 6-bit Shift register to store the $PRNG$ seed, (iii) a 6-bit primitive LFSR built of 6 flip-flops and 1 XOR gate, and (iv) one 64-bit MUX and a 1-bit De-MUX. The $Scrambler$ is implemented in software as a simple algorithm that repositions the challenge bits based on the LFSR's outputs in consecutive clock cycles. Finally, a small controller, implemented as a 6-state finite state machine, manages DAUP. 
Moreover, for this 64-bit PUF, 6 clock cycles are needed to initialize the PRNG and 63 clock cycles to generate the scrambling patterns. Such latency for an IoT device operating at 1GHz is in the order of 100ns, which is quite insignificant compared to the time needed to transfer/encode/decode packets (order of ms).
Similarly, the added hardware will draw negligible power.  
Finally, we evaluated the impact of measurement noise in our setup by repeating each experiment. In $~\approx$1\% of the cases we observed noisy responses when DAUP is used.
Error correction codes (ECC) and majority voting over multiple measurements can be used to mitigate this negligible noise~\cite{Chatterjee-19-Building}.



\vspace{-0.05cm}
\section{Conclusion}\label{sec:conclusion}
IoT is characterized by the large-scale involvement of nodes and spatial coverage.
Therefore accessibility and availability of a server for all IoT devices is not always guaranteed. This makes distributed authentication plausible. In this paper, we have proposed DAUP, a low overhead and highly secure distributed PUF-based authentication scheme that fits the resource-constrained IoT devices. DAUP leverages PUFs as device fingerprints and employs challenge scrambling to safeguard them against modeling attacks. The scrambling pattern depends on the queried challenge, versifier's ID, and prover's unclonable PUF signature. The validation results confirm the efficacy of the proposed method against contemporary security attacks, both PUF modeling and  conventional attacks launched in IoT frameworks such as impersonation and Sybil attacks. 

\vspace{-0.1cm}
\section*{Acknowledgement}
This work has been  supported by the National Science Foundation MRI Award (1920079).

\vspace{-0.05cm}
\scriptsize
\bibliographystyle{IEEEtran}
\bibliography{main.bib}

\end{document}